\documentclass[twocolumn,showpacs,amssymb,nobibnotes,aps,floats,prb]{revtex4-1}
\usepackage{textcomp,amssymb,graphicx}
\usepackage{hyperref}
\usepackage{amsbsy}
\usepackage{amsmath}
\usepackage[dvips]{color}
\usepackage{graphicx}
\usepackage{float}
\date{}

\begin{document}
\title{Skyrmion crystal phases in antiferromagnetic itinerant triangular magnets}
\author{Sahinur Reja}
\affiliation{Department of Physics, School of Physical Sciences, Central University of Rajasthan, Bandasrsindri, Kishangarh-305817, Rajasthan, India }
\date{\today}
\begin{abstract}
{Very often the skyrmions form a triangular crystal in chiral magnets. Here we study the effect of itinerant electrons 
on the structure of skyrmion crystal (SkX) on triangular lattice using Kondo lattice model in the large coupling limit and treating the localized spins as classical vectors. To simulate the system, we employ
hybrid Markov Chain Monte Carlo method (hMCMC) which includes electron diagonalization in each MCMC update for classical spins. We present the low temperature results for $12\times 12$ system at electron density $n=1/3$ which show a sudden jump in skyrmion number when we increase the hopping strength of the itinerant electrons. We find that this high skyrmion number SkX phase is stabilized by combined effects: lowering of density of states at electron filling $n=1/3$ and also pushing the bottom energy states further down. We show that these results hold for larger system using travelling cluster variation of hMCMC. We expect that itinerant triangular magnets might exhibit the possible transition between low density to high density SkX phases  by applying external pressure.}
\end{abstract}
\maketitle
\section{Introduction}
Magnetic Skyrmions are topologically protected local whirls of the spin configuration recenly observed in non-centrosymmetric magnetic materials\cite{skyrmion1,skyrmion2}, and thin films \cite{skyrmion_thin_film}. These nanoscale spin textures are usually induced by chiral interaction of the Dzyaloshinskii–Moriya (DM) type\cite{Dzyaloshinskii, Moriya} and can be characterised  by the topological number often called Skyrmion number which measures the winding of the normalized local magnetization. Topological protection of magnetic skyrmion attracts scientists due to its potential applications in information processing and computing\cite{skyrmion_thin_film,albert_fert}
In particular, isolated {\it mobile} magnetic skyrmions working as data carriers might in principle be channelled to mechanically fixed reading/writing device, for instance by applying
an electric field in a race-track memory setup.\cite{race_track_Parkin,race_track_Bauer} But it is not easy to move the skyrmions along the current due to the skyrmion Hall effect\cite{skyrmion_hall} induced by its nonzero topological number.  

Not only in ferromagnets, but the skyrmions are found in antiferromagnatic materials where Skyrmion are formed as a pair of strongly coupled
topological objects corresponding to each sublattice. So, unlike the FM counterpart, the Magnus force is cancelled by the opposing topological index of each sublattice. The absence of Magnus force might be useful in current induced movement of skyrmions\cite{afm_skyrmion1,afm_skyrmion2} in spintronics applications. Also, the DM interaction which is essential in stabilizing the skyrmion is more common in AF materials.

\begin{figure}[htb]
\centering
\includegraphics[width=.7\linewidth,angle=0]{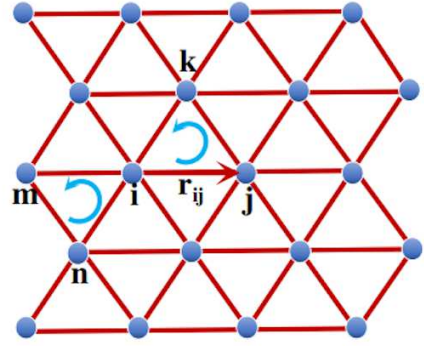}
\caption{(color online) Schematic of the model on triangular lattice: unit vectors {\bf $\hat{r}_{ij}$} connecting sites $i$ to $j$ are used to define the DM interactions between spin $\bf S_i$ and $S_j$. The indices $i, j, k...$ etc are used in anti-clockwise direction to calculate the skyrmion number defined in eq. \ref{eq_skyrmion_num}}
\label{fig_tri_latt}
\end{figure}

However, it is often found that the skyrmions arrange themselves in triangular lattice structure called Skyrmion crystal (SkX) which has been experimentally observed in metallic ferromagnet MnSi\cite{skyrmion1}. The noncoplanar
spin texture in the SkX is described by a superposition of
three non-equivalent helical spin density waves\cite{skyrmion1,skyrmion2,rossler}, which often result due to the competition between the ferromagnetic exchange and the DM interaction originating from the spin-orbit coupling. In square lattice, the SkX is stabilized by the competition of DM interaction and ferromagnetic interaction along with the applied magnetic field and thermal fluctuations. But not only FM interactions, but also AFM interaction along with DM interactions are shown to stabilize SkX when a magnetic field is applied. For instance, 
a triangular lattice isotropic Heisenberg model with strong FM near-neighbour and weak AFM interaction has been shown to stabilize the SkX\cite{okubo_triangular}. More recently, it has been shown that  pure AFM classical Heisenberg spin model on frustrated triangular lattice in a magnetic field and a small DM interaction can stabilize SkX spin texture. This antiferomagnetic skyrmion crystal (AF-SkX) consists of three interpenetrating Skyrmion lattices (one by sublattice) SkX which survives upto very low temperature\cite{three_sub_tri_SkX}. 

Most of the theoretical studies on skyrmions consider spin only models.  In this paper, we intend to study the effect of itinerant electrons on the the structure of SkX formed on frustrated lattice, e.g., on triangular lattice described above. This active field of research is interesting because the conduction electrons coupled to localized spins give rise to exotic multiple-Q magnetic orders\cite{multiple_Q1,multiple_Q2,multiple_Q3,multiple_Q4}, coupled spin-charge phases\cite{spin_charge_order1,sahinur_sanjeev1,spin_charge_order2}. Interestingly, the present author and collaborators find an electronic route to stabilize nanoscale spin texture in a triangular lattice model with low density conduction electrons Kondo coupled to localized magnetic moments\cite{sahinur_sanjeev2}. Also, very recently, a SkX with an unusually higher topological number of two is
found to be stabilized in itinerant triangular lattice Kondo model with longer range electron hopping \cite{skyrmion_number2}.

In this paper, we study the effect of itinerant electrons on the structure of AF-SkX\cite{three_sub_tri_SkX} triangular lattice model with classical spins using Kondo-Lattice model. 
To study the ground state properties, we employ
hybrid Markov Chain Monte Carlo (hMCMC) which includes electron diagonalization in each Monte Carlo update for classical spins. For electron density $n=1/3$, we find that the skyrmion number abruptly jumps to higher value and slowly goes to zero when we increase the electron hopping parameter. This is evident also from the real space MCMC spin configuration for AF-SkX. The higher skyrmion density is favoured because the system gains energy by the reduction of electron density of states at $n=1/3$ and pushing the electronic states down. 

The paper is organized as follows. We give the formal definition of the model and methods we use to simulate the system in Sec 2. In sec 3, the results are presented and finally conclude in sec 4.

\begin{figure}[htb]
\centering
\includegraphics[width=.7\linewidth,angle=-90]{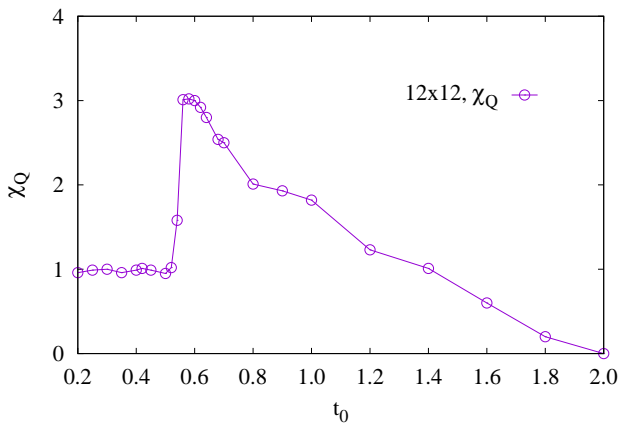}
\caption{(color online) Skyrmion number as a funtion of electron hopping strength $t_0$ with $n=1/3$ electron filling. The other parameters are $J=0.25, D=0.125, B=0.7$}.
\label{fig_skyrmion_number_12x12}
\end{figure}

\section{Model and method}
We consider a spin model on triangular lattice on which a certain parameter regime stabilizes the skyrmion crystal (SkX). The Hamiltonian reads,
\begin{eqnarray}
H_m=J\sum_{\langle ij\rangle}{\bf S_i\cdot S_j}+D\sum_{\langle ij\rangle}{\bf \hat{r}_{ij}}\cdot {\bf S_i\times S_j}-B\sum_iS_i^z
\label{eq_class_model}
\end{eqnarray}
where ${\bf S_i}$ is the unimodular classical spin variable at site $i$ on triangular lattice as shown in Fig.\ref{tri_latt.eps}. The first and last term in Eq.\ref{eq_class_model} represent the classical Heisenberg interaction and external magnetic field term. The 2nd term is DM coupling between two sites $i$ and $j$ with ${\bf \hat{r}_{ij}}$ being a unit vector pointing from $i$ to $j$. Also, ${\langle ij\rangle}$ in the summation indicates nearest-neighbour (NN) interaction.
This classical spin model is known to give rise to SkX for certain parameter range and even at finite temperature\cite{three_sub_tri_SkX}.

To study the effect of itinerant electrons on the structure of SkX, we couple each classical spins in $H_m$ with the quantum spin $s_i$ of conduction electrons, just as in Kondo lattice model. So the total Hamiltonian we would be considering,  
\begin{eqnarray}
H=H_m+\sum_{\langle ij\rangle\sigma}(t_{ij}c_{i\sigma}^{\dagger}c_{j\sigma}+h.c)+J_k\sum_{i}{\bf S_i\cdot s_i}
\label{eq_total_model}
\end{eqnarray}
Note that the localized classical spin at a site $i$ tries to  align the electron spins at that site, giving the spin up and down conduction bands separated by Kondo coupling $J_k$.  So, in large coupling limit, only one type of electrons contribute to the low energy physics i.e., we can omit the spin index $\sigma$ and the Eq.\ref{eq_total_model} becomes,
\begin{eqnarray}
H=H_m+\sum_{\langle ij\rangle}(t_{ij}c_{i}^{\dagger}c_{j}+h.c)
\label{eq_model_simulation}
\end{eqnarray}
where the hopping of conduction electrons gets modified as $t_{ij}=t_0[cos(\theta_i/2)cos(\theta_j/2+sin(\theta_i/2)sin(\theta_j/2)e^{i(\phi_i-\phi_j)}]$ from site $i$ to $j$ which now depends on the orientation of localized spins at site $i$ and $j$ defined by polar and azimuthal angle $(\theta_i,\phi_i)$ and $(\theta_j,\phi_j)$ respectively.  

We employ hMCMC method to simulate the systems. This includes MCMC method for localized spins which are treated as classical unit vector. But for each MCMC updates of spin orientation, we need to calculate the electronic energy by diagonalizing the $t_{ij}$ Hamiltonian matrix and filling the energy states upto the number of electrons defined by a given density $n$. We present the results for $n=1/3$, i.e., one electron per three sites. The time consuming diagonalization of $t_{ij}$ matrix does not allow the simulation for larger systems as we are usually capable of handling in MCMC simulations. So, we first present the results on a smaller $12\times 12$ system which captures the full characteristics of the results. We use simulated annealing process i.e., our simulation starts at high temperatures and slowly going down to very low temperature $T=0.001$ for which we present the results. Typically we use $10^3$ MCMC steps for equilibration at each temperature and similar number for measurements of physical quantities. 

We also present the results on larger systems on $N\times N=24\times 24$ using a variant of hMCMC method called travelling cluster approximation (TCA)\cite{TCA_sanjeev}. The  basic idea relies on the approximation that a MCMC update for a spin at site $i$ affects only the elements around that site in $t_{ij}$ hopping matrix. So, it is sensible to diagonalize $t_{ij}$ matrix for a cluster $N_c\times N_c$ (with $N_c\leq N$) around site $i$ to calculate the electronic energy contribution, rather than diagonalizing the Hamiltonian $t_{ij}$ for whole system $N\times N$. Only thing we need to remember that the electronic density for the cluster $n_c=n(N_c\times N_c)/(N\times N)$. The cluster size could be increased to get the more accurate results and of course cluster size $N_c\times N_c=N\times N$ gives the exact hMCMC results. 
We used cluster sizes $6\times 6$ and  $12\time 12$ for simulations in $24\times 24$ systems and notice a negligible finite size effect.

\begin{figure}[h!]
\centering
\includegraphics[width=.8\linewidth,angle=-90]{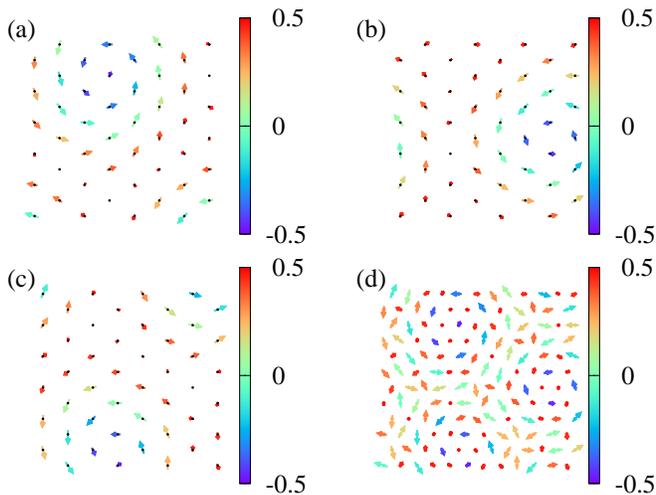}
\caption{(color online) The real space spin configurations from hMCMC simulation at $t_0=0.5$: (a), (b) and (c) represent the spin orientations for one of the three sublattice structures, whereas (d) gives the combined spin configurations. The parameters for $H_m$ are $J=0.25, D=0.125, B=0.7$. The color bar measures the z-components of spins.}
\label{fig_spin_config_t_0.5}
\end{figure}

\section{Results}
As mentioned above, we study the spin Hamiltonian in \ref{eq_class_model} on triangular lattice with localized spins interacting with conduction electrons. This spin only model at low temperature is known to give three sublattice SkX phase for $J=0.25, D=0.125$ and $B\sim 0.5$ to $ \sim 1.6$ when spins are treated as classical vectors\cite{three_sub_tri_SkX}. We chose the parameters $J=0.25, D=0.125, B=0.7$ to be in the SkX phase and add the itinerant electrons Kondo coupled to these classical spins to study it's effect on the SkX phase by changing the electron hopping strength $t_0$.  

\begin{figure}[htb]
\centering
\includegraphics[width=.8\linewidth,angle=-90]{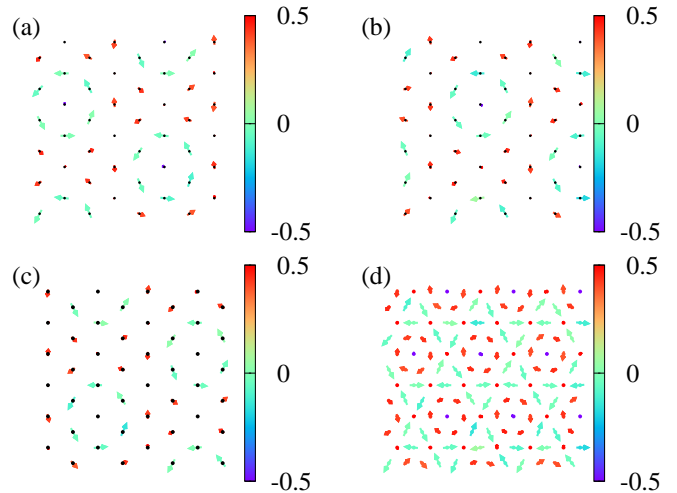}
\caption{(color online) The similar results for spin orientations as in Fig.\ref{fig_spin_config_t_0.5}, but for larger hopping strength $t_0=0.6$.}
\label{fig_spin_config_t_0.6}
\end{figure}

\subsubsection{Results for $12\times 12$ system}
The simulation after adding the electrons to the system becomes time consuming as compared to the usual MCMC method for spin only systems. This is because we now need to diagonalize the electronic Hamiltonian in each MCMC update for spin orientations. So, larger system simulation is almost impossible in this method of hMCMC. Here we choose the system size to be $12\times 12$ to capture the basic results.  For electron density $n=1/3$, we run the simulation for different electron hopping parameter $t_0$ and keep track of how the SkX phase changes. To identify the changes, we calculate the topological charge i.e., skyrmion number by defining its discretized version as\cite{three_sub_tri_SkX}:

\begin{eqnarray}
\chi_Q=\frac{1}{4\pi}\left\langle\sum_i A_i^{jk}{\rm sign}\left(\chi_i^{jk}\right)+A_i^{mn}{\rm sign}\left(\chi_i^{mn}\right)\right\rangle
\label{eq_skyrmion_num}
\end{eqnarray}
where  $\chi_i^{jk}={\bf S_i.S_j{\times}S_k}$ is the local scaler chirality which measures the volume created by three spins at site $i, j, k$ as indicated in Fig.\ref{fig_tri_latt}. Also, $A_i^{jk}=||{\bf (S_j-S_i){\rm x}(S_k-S_i)}||/2$ is the local area of the surface spanned by three spins at site $i, j, k$. Here $\langle\dots\rangle$ represents the MCMC average after thermal equilibration. As we mentioned earlier, the AF-SkX phase on triangular lattice forms on each of the three sublattices. So, we compute the Skyrmion number on each sublattice i.e., the sites $i, j, k$ run over one sublattice only. 

Fig.\ref{fig_skyrmion_number_12x12} shows how skyrmion number for each sublattice varies in $12\times 12$ lattice system when we change the electron hopping term $t_0$ at electron density $n=1/3$ and with paramters $J=0.25, D=0.125, B=0.7$. For smaller $t_0$, the electrons seem to have very small effect on SkX. So, we notice that skyrmion number remains constant to 1 which is evident from the real space spin configuration at $t_0=0.5$  as shown in Fig.\ref{fig_spin_config_t_0.5}(a),(b),(c) and (d) respectively for three different sublattices and all together. But, with increasing $t_0$, the effect of electrons is to polarising the localized spins as the electron hopping becomes the easiest if the neighbouring localized spins are parallel. This process of electron mediated ferromagnetic interaction between the localized spins is called {\it double exchange} mechanism. This effect at large $t_0$ gives a FM spin configuration and eventually destroys the SkX phase to give the skyrmion number gradually going to zero as shown in Fig.\ref{fig_skyrmion_number_12x12}. But in the intermediate $t_0$ values, the itinerant electrons stabilize a SkX phase with larger skyrmion number. The real space spin configuration obtained from hMCMC simulations at $t_0=0.6$ in this phase for different sublattices and all together is shown in Fig.\ref{fig_spin_config_t_0.5}(a),(b),(c) and (d) respectively. The transition from lower to higher skyrmion number phase seems to be abrupt and then the skyrmion number gradually decreases to zero as $t_0$ is increased.  

\begin{figure}[htb]
\centering
\includegraphics[width=.7\linewidth,angle=-90]{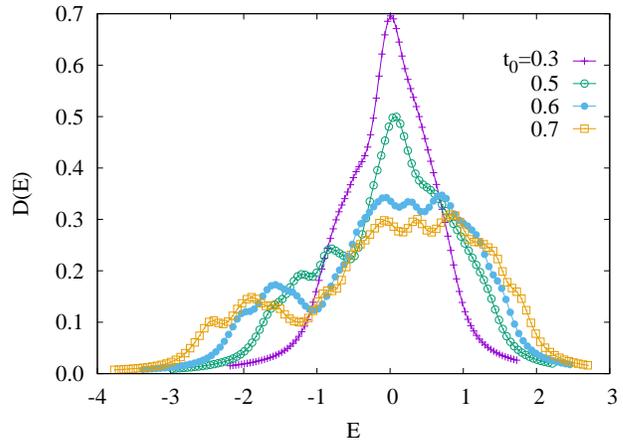}
\caption{(color online) The electronic density of states for different values of $t_0$. The other parameters are $J=0.25, D=0.125, B=0.7$.}
\label{fig_dos_12x12}
\end{figure}

To understand this better, we looked at the density of states $D(E)$ for different values of $t_0$ shown in Fig.\ref{fig_dos_12x12}. For smaller $t_0$, $D(E)$ forms a bit narrow electronic band. As we increase $t_0$, we notice that the lower energy states get pushed down further and there is a reduction (more prominent for $24\time 24$ system as shown below) of density of states corresponding to $n=1/3$. In this configuration of higher skyrmion number the electron gains more energy and this helps lowering the total energy of the system. So, the larger skyrmion density spin configuration gets stabilized at intermediate $t_0$.   

\begin{figure}[htb]
\centering
\includegraphics[width=.7\linewidth,angle=-90]{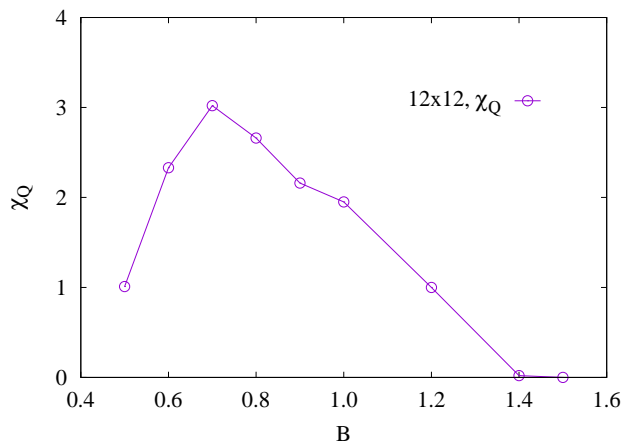}
\caption{(color online) The skyrmion number for different values of $B$ at $t_0=0.6$. The other parameters are $J=0.25, D=0.125$.}
\label{fig_skyrmion_number_vs_B_12x12}
\end{figure}

We note that both the electrons hopping and magnetic field have the similar spin polarising effect; thus decreasing the skyrmion number and eventually destroying the SkX phase at large $t_0$ and larger $B$ respectively. So, we study the effect of $B$ on the SkX phase at $t_0=0.6$ with larger skyrmion number. We find that the skyrmion number gradually decreases with increasing and decreasing $B$ with maximum skyrmion number at $B\sim 0.7$ as shown in Fig.\ref{fig_skyrmion_number_vs_B_12x12}. This is why we choose to show the above results for $B=0.7$.     

\begin{figure}[htb]
\centering
\includegraphics[width=.46\linewidth,angle=-90]{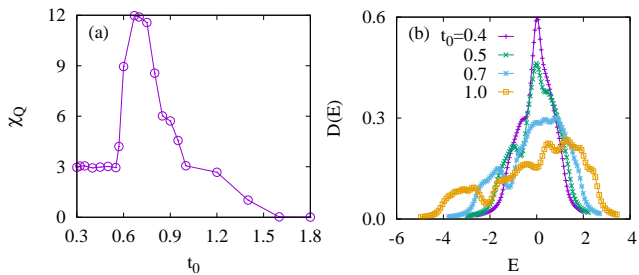}
\caption{(color online) The skyrmion number and density of states for different $t_0$. The other parameters are $B=0.7, J=0.25, D=0.125$.}
\label{fig_skyrmion_dos_24x24}
\end{figure}

\begin{figure}[h!]
\centering
\includegraphics[width=.8\linewidth,angle=-90]{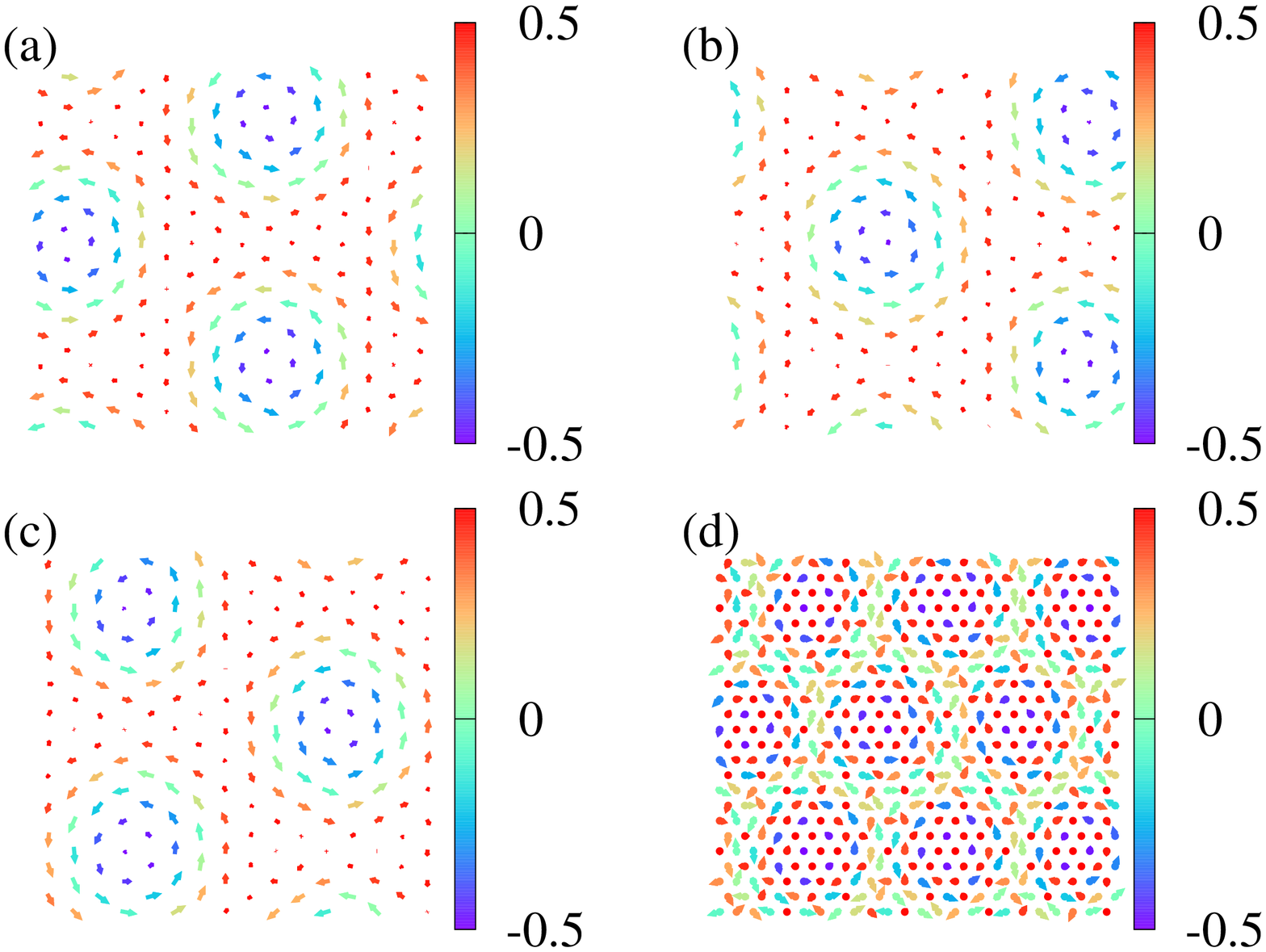}
\caption{(color online) The real space spin configurations from MCMC simulation at $t_0=0.3$: (a), (b) and (c) represent the spin orientations for one of the three sublattice structure, whereas (d) gives the combined spin configurations. The parameters for $H_m$ are $J=0.25, D=0.125, B=0.7$. The color bar measures the z-components of spins.}
\label{fig_spin_config_24x24_t0_0.3}
\end{figure}

\begin{figure}[htb]
\centering
\includegraphics[width=.8\linewidth,angle=-90]{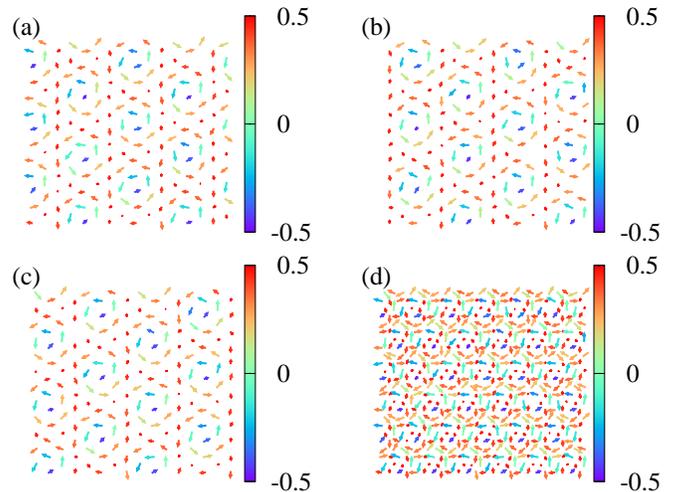}
\caption{(color online) The similar results for spin orientations as in Fig.\ref{fig_spin_config_24x24_t0_0.3}, but for larger hopping $t_0=0.7$.}
\label{fig_spin_config_24x24_t0_0.7}
\end{figure}

\subsubsection{Results for $24\times 24$ system}
To substantiate the findings, we now present the results for $24\times 24$ system. This simulations is done by using a variant of
hMCMC method called TCA (mentioned above)\cite{TCA_sanjeev}. We present the simulation results by taking cluster size to be $6\times 6$. This has been checked to give same results when we choose the cluster size $12\times 12$. As we mentioned earlier that the main time consuming part of this kind of simulations is the diagonalization of electronic Hamiltonian. So, we restrict us to limited number of measurements ($\sim 100$) which needs to diagonalize the  $576\times 576$ matrix Hamiltonian for electrons which is the $t_{ij}$ matrix size on $24\times 24$ system.

Fig.\ref{fig_skyrmion_dos_24x24}(a) and (b) respectively show the skyrmion number and density of states for different $t_0$. Again, we choose the suitable magnetic field $B=0.7$ which gives the maximum skyrmion number. We notice the similar behaviour of skyrmion number variation with hopping strength $t_0$ as in the case of $12\times 12$ systems. Smaller $t_0$ does not affect the skyrmion structure with about three skyrmions per sublattice which is consistent with the results for $12\times 12$ systems. But with increasing $t_0$ further, the skyrmion number jumps almost suddenly to higher values and gradually decreases to zero due to the polarising effect of electrons on localized spins. Looking at the density of states in (b), we see again the higher skyrmion density is stabilized because it gains the electronic energy by pushing the electronic states down as well as the reducing $D(E)$ at $n=1/3$ filling. The real space spin configurations for $t_0=0.3$ and $t_0=0.7$ are shown in Fig.\ref{fig_spin_config_24x24_t0_0.3} and Fig.\ref{fig_spin_config_24x24_t0_0.7} respectively.

\section{Conclusions}
In conclusion, we study the effect of the itinerant electrons on the skyrmion crystal (SkX) phase on triangular lattice using Kondo lattice model in the large Kondo coupling limit. We employ hybrid MCMC method which treats the localized spins classically, but the electronic Hamiltonian gets diagonalized to calculate the kinetic energy of the system for each MCMC step; thus limiting the calculations to smaller system sizes. We start our simulation at higher temperature and slowly go to lowest temperature ($T\sim 0.001$) with about $10^3$ MCMC steps for each temperature for equilibration of the system and similar number of measurement has been carried out. 

Keeping the parameters $J=0.25, D=0.125, B=0.7$ for magnetic Hamiltonian, we simulate a triangular lattice  $12\times 12$ system with electron density $n=1/3$ and for different values of electron hopping strength $t_0$. Smaller $t_0$ does not affect the SkX phase with about one skyrmion per sublattice found in spin only model\cite{three_sub_tri_SkX}. But with increasing $t_0$, the skyrmion number abruptly jumps to three per sublattice which are consistent with the real space spin configurations found in hMCMC simulations. We show that the higher skyrmion density phase is stabilized due to combined effect: reduction of electronic density of states at about $n=1/3$ and pushing the bottom electron energy states further down. With larger $t_0$, the electrons gain kinetic energy by mediating the ferromagnetic interaction ({\it double exchange mechanism}) between the localized spins. Thus we notice the skyrmion number slowly decreases after the higher skyrmion density phase and finally goes to zero at large $t_0$. Using a variant of
hMCMC method called TCA\cite{TCA_sanjeev}, we show that these results hold for larger $24\times 24$ systems. We speculate that external pressure on itinerant triangular magnets might show the possible transition between low density to high density skyrmion crystal phases.

\section{Acknowledgments}
We would like to thank U. Nitzsche for technical assistance.  Computations were
carried out on the ITF compute cluster in IFW Dresden, Germany.

\bibliography{skyrmion,sahinur_paper}
\end{document}